\documentclass[journal]{IEEEtran}

\usepackage[utf8]{inputenc}
\usepackage[T1]{fontenc}
\usepackage{array}
\usepackage{amsmath}
\usepackage{colortbl}
\usepackage[table*]{xcolor}
\usepackage{svg}
\usepackage{subcaption}
\usepackage{multirow}
\usepackage{float}
\usepackage{lipsum}
\usepackage{xcolor}
\usepackage[export]{adjustbox}
\usepackage{comment}
\usepackage{xcolor}
\usepackage{bm}
\usepackage{cite}
\usepackage{soul}
\usepackage{subcaption}
\usepackage{sidecap}
\usepackage{enumitem}
\usepackage{booktabs}
\xdefinecolor{gray95}{gray}{0.65}
\xdefinecolor{gray25}{gray}{0.75}
\usepackage{lipsum}
\usepackage{ulem}
\usepackage{rotating}
\usepackage[ruled,linesnumbered]{algorithm2e}
\begin{document}
\title{Exploring and Enhancing Placement of IDS in RPL:\\ A Federated Learning-based Approach}

%
%
%

\author{Selim Yılmaz,
        Sevil Sen,
        and Emre Aydogan
\thanks{S. Yılmaz, E. Aydogan, and S. Sen are with the WISE Lab., Department of Computer Engineering, Hacettepe University, Ankara 06800, Türkiye e-mail: (selimyilmaz@mu.edu.tr, ssen@cs.hacettepe.edu.tr, emre.aydogan@karel.com.tr).}
\thanks{S. Yılmaz is also with the Department of Software Engineering, Muğla Sıtkı Koçman University, Muğla 48000, Türkiye.}
\thanks{E. Aydogan is also with Karel Electronics, Ankara 06800, Türkiye}
}

\maketitle

\begin{abstract}
In RPL security, intrusion detection (ID) plays a vital role, especially given its susceptibility to attacks, particularly those carried out by insider threats. While numerous studies in the literature have proposed intrusion detection systems (IDS) utilizing diverse techniques, the placement of such systems within RPL topology remains largely unexplored. This study aims to address this gap by rigorously evaluating three intrusion detection architectures, considering central and distributed placement, across multiple criteria including effectiveness, cost, privacy, and security. The findings underscore the significant impact of attacker position and the proximity of IDS to attackers on detection outcomes. Hence, alongside the evaluation of traditional intrusion detection architectures, this study explores the use of federated learning (FL) for improving intrusion detection within RPL networks. FL's decentralized model training approach effectively addresses the impact of attacker position on IDS performance by ensuring the collection of relevant information from nodes regardless of their proximity to potential attackers. Moreover, this approach not only mitigates security concerns but also minimizes communication overhead among ID nodes. Consequently, FL reduces the need for extensive data transfer, thus mitigating the impact of packet loss and latency inherent in lossy networks. Additionally, the study investigates the effect of local data sharing on FL performance, clarifying the balance between effectiveness and security.
\end{abstract}


\section{Introduction}
\label{section_introduction}

\IEEEPARstart{I}{nternet} of Things becomes pervasive in many applications from industrial IoT to the Internet of medical things, the Internet of drones. While there are 13.1 billion IoT connected devices worldwide as in 2022, this number is predicted to reach up to 29.4 billion devices by 2030 \cite{Statista}. The low power and lossy networks (LLN) are a special type of IoT that takes place in various application areas such as industry, smart homes. In such networks, resource constrained devices are connected over lossy links, which results in high packet loss. Therefore, new routing protocols are proposed for providing communication in such lossy networks in the literature. Among them, routing protocol for low power and lossy networks (RPL) has become the standard routing protocol for LLNs. However, RPL is still open to improvements in many research areas such as load balancing, mobility, security. 

In the last years, a lot of approaches are proposed for securing RPL. While some specifications are defined for external attacks against RPL in its RFC document \cite{rfc6550}, it is still vulnerable to insider attacks. Therefore, researchers have been working on analyzing RPL-specific attacks and developing IDSs for  RPL in order to respond to such attacks in a timely manner. However, the position of attackers and the placement of IDSs could dramatically affect the performance of the proposed IDSs. While studies primarily concentrate on improving intrusion detection methods, the placement of IDSs significantly influences the efficacy of any proposal. Therefore, this study initially conducted a comprehensive exploration of IDS placement in response to various types of RPL attacks launched by attackers situated in different locations. Subsequently, we introduce an IDS solution, which utilizes federated learning (FL), aiming to overcome the primary shortcomings of current architectures in terms of both effectiveness and efficiency.

In order to accomplish our first goal, we take three different architectures into consideration: \textit{i}) central ID with local information (CIDwL), \textit{ii}) central ID with global information (CIDwG), and \textit{iii}) distributed and collaborative ID (DCID). In CIDwL, IDS is placed at one node which relies on its local information in order to reach decisions. Many studies in the literature rely on a central IDS placed at the root node due to being a data collector point and a more powerful device in terms of computation capabilities \cite{PRISM}. In CIDwG, the central IDS collects information from other nodes in order to have enough data for reaching to a decision. Lastly, DCID, in which every node has an IDS agents and decides upon collaboratively by using a voting mechanism, is analyzed.

This study provides a comprehensive analysis of IDS placement, an aspect often overlooked in existing research on intrusion detection. Our analysis demonstrates the significant impact of IDS placement on IDS performance, comparing three placement strategies (CIDwL, CIDwG, DCID) across various simulation scenarios (31,626 scenarios). To the best of our knowledge, this is the first  study that analyze the placement of intrusion detection system in RPL topology comprehensively from different aspects that are accuracy, cost, response time, privacy, and security. The insights from this analysis can serve as valuable guidance for future research efforts aiming to propose IDS solutions for RPL, guiding their evaluation processes effectively. Building on this guidance, our study also proposes a novel IDS designed to balance effectiveness and efficiency.

Here, a rigorous analysis is carried out for answering the following research questions:

\noindent\textbf{RQ1}: Is one central IDS enough for effectively detecting all types of attacks that are performed at different locations? 

\noindent\textbf{RQ2}: What is the minimum number of IDS required for effective detection? 

\noindent\textbf{RQ3}: How do IDSs make decision together efficiently from the communication cost perspective? 

The results show that the position of ID nodes has a significant impact on IDS performance, depending on the attack type and attacker location. A single central ID node, such as one placed at the root node, proves ineffective against attacks from multiple positions. Hence, distributing ID nodes emerges as a promising alternative, particularly by placing at least one ID node in each level. This approach enables each ID node to effectively detect attacks in its proximity. However, the addition of extra ID nodes may lead to increased overhead and energy consumption, consequently reducing the network's overall lifetime. Furthermore, communication challenges arise due to lossy links between IDS agents. Thus, deploying an IDS that balances both efficacy and efficiency factors becomes imperative, especially in LLNs.

In this study, we also introduce a federated learning-based intrusion detection (FedID) system, which is another goal of this study. FedID incorporates collaborator nodes in the training process, where instead of directly sharing data, each collaborator shares the weights of their local model. This sharing of weights enables the generation of a global model representative of the collective knowledge of all collaborator nodes and perform better detection. Additionally, weight sharing ensures that communication remains private and reduces the energy consumption within the system, as it eliminates the need for transmitting large amounts of data. However, in this study, each collaborator also shares a small portion of local traffic samples (up to 10\%) to enhance the effectiveness of the global model. As shown experimentally, this approach is necessary due to the non-Independently and Identically Distributed (non-IID) nature of the data, resulting from differences in local traffic patterns caused by attackers' locations.
The experimental results show that FedID offers superior detection performance than CIDwL, competitive performance with much lower cost, particularly after model's development, in comparison to CIDwG and DCID, making it highly suitable for LLNs.

The paper is organized as follows. Section~\ref{section_background} covers background information, including RPL and attacks. Section~\ref{section_related-work} discusses related studies in the field of intrusion detection on RPL. In Section~\ref{section_architecture-analysis}, we introduce and analyze various ID architectures in detail. The key findings observed from the analysis of architectures are given and discussed in detail in Section~\ref{section_comparative-analysis-architecture}. Section~\ref{section_federated-architecture} presents FedID, offering a comparative evaluation of its detection performance and discussing its advantages and disadvantages over other architectures.
Section~\ref{section_discussion} provides a comparative discussion of the results for each architecture, including FedID, and outlines possible future studies. Finally, Section~\ref{section_conclusion} concludes the findings of this study.

\section{Background}
\label{section_background}

\subsection{Protocol Overview}
RPL is a distance vector routing protocol in which nodes communicate over a special topology called destination-oriented directed acyclic graph (DODAG). The formation of DODAG is initiated by the root node only by broadcasting control packets called DODAG information object (DIO). The DIO package carries the information needed for the nodes to join the DODAG. Each node that receives the DIO packet adds the sender address in the incoming DIO packet to its parent list. In addition, each node calculates `rank' value according to the objective function (OF) defined in DIO to determine its position in DODAG with respect to the root node. This guarantees the acyclic nature of the graph. After the DIO packet is updated with its own address and rank value, the node forwards it to its neighbors. In this way, an \textit{upward route} is established from the leaf nodes to the root node. 

A protocol-specific algorithm called trickle timer controls the transmission interval of DIO packets. DIO packets are initially broadcast more often for the fast attachment of nodes to the graph, and the interval increases as long as the network is stable. A new node does not necessarily wait for a DIO packet to join DODAG; instead, it sends a control package called DODAG information solicitation (DIS) to its neighbors. The neighboring node sends the DIO packet immediately after taking the DIS packet. Therefore, the new node can join the DODAG. The \textit{downward route} in RPL, however, is realized with control packages called destination advertisement object (DAO). The downward routes are built in two modes: \textit{storage mode} and \textit{non-storing mode}. In the storage mode, each node keeps a routing table, and, instead of sending them towards the root node, it forwards packets to the next hop that routes to the destination address. In non-storage mode, however, the routing table is kept by the root node only, and the packets must be forwarded to the root node which operates the downward routing.

\subsection{RPL Attacks}
RPL control messages can be exploited to change the topology of a network. Furthermore, IoT nodes are more susceptible to various attacks due to mobility and their resource constraints. In the literature, RPL attacks are split into three categories~\cite{mayzaud2016taxonomy}; attacks on topology, attacks on resources, and attacks on traffic. The first category consists of two subcategories; sub-optimization and isolation. The former creates non-optimal routes that cause poor performance in the network, and the latter deals with separation of a node or nodes from the network so that they cannot communicate with other nodes in the network. Attacks on resources cause an increase in computing, communication, energy usage in nodes and can affect local or whole network. Attacks on traffic, however, aims to decept the legitimate nodes by claiming other node's identity, as well as passive listening them by focusing on eavesdropping activities. In this study, we focus on the following seven RPL attacks belonging to different categories.

\begin{itemize}
    \item \textit{Decreased Rank (DR):} The attacker nodes decrease the value of its rank and send it to other nodes in the network. Due to lower ranks denote higher position in DODAG tree and being closer to the root node, attacker nodes are preferred as parent nodes so that a large portion of traffic goes through these attacker nodes. 
    \item \textit{Increased Version (IV):} Only the root node in DODAG tree is responsible for increasing the version number of DODAG tree and advertising version number throughout the network for global repair in RPL normally. However, attacker nodes illegitimately increase version number and propagate it, which may cause unnecessarily rebuilding of the networks. 
    \item \textit{Blackhole (BH):} In this denial-of-service attack, instead of forwarding incoming packets to neighboring nodes, attacker nodes drop all packets. If it is combined with a sinkhole attack, attackers can damage the network in a way that a large portion of the network traffic is lost.
    \item \textit{Selective Forwarding (SF):} While blackhole attack drops all packets that are supposed to be forward, selective forwarding randomly selects some specific packets and drops them. Hence, packets belonging to a particular protocol can be filtered, and routing paths can be disrupted. 
    \item \textit{Worst Parent (WP):} A node chooses a parent node when new incoming packets to be forwarded come according to the OF. In this attack, the attacker nodes exploit the objective function and force the victim node to choose the worst possible parent to forward packets. As a result, unoptimized paths are constructed, causing poor network performance.
    \item \textit{DAG Inconsistency (DI):} Downward route are built by DAO messages in RPL. If a child node cannot send packets to the destination due to unavailable downward routes which are built as a result of fake DAO messages, an inconsistency is occurred. When a node receives a packet with the Forwarding-Error `F' flag set, this indicates that the packet cannot be sent by a child node and sent back to the parent node to choose a new neighbor node. As a result, node or nodes can be separated from network and unoptimized topology is emerged.
    \item \textit{Hello Flood (HF):} A node broadcasts an initial message when trying to join the network. It is called a HELLO message and is sent to the node's neighbors within its communication range. Attacker exploits this mechanism and regularly sends a good amount of unnecessary HELLO messages to its neighbors. The victim node replies with DIO messages to these HELLO messages and resets its trickle timer. This attack increases control packet overhead and node energy consumption.

\end{itemize}

\section{Related Work}
\label{section_related-work}
There has been an increase in studies on RPL in recent years, but here we specifically review studies from an architectural perspective. Following this, the discussion extends to examining the applications of federated learning  in the context of IDS for IoT, illustrating its potential to enhance security in this domain.

Early studies generally adopted centralized placement of IDSs where a node, usually a sink/root or LLN border router (LBR) node, is chosen as an ID node. In \cite{7921038,8276832,OpinionMetric}, trust-based systems in which the monitoring nodes passively collect and send information to the sink node, which is responsible for analyzing the incoming information. In addition, machine learning-based intelligent systems are also used~\cite{ML-IDS,Foley2020,8270652,8777504,10.1145/2990499, MI-BGSA,TL-GP}. K-means clustering and decision tree-based supervised learning approaches are employed in~\cite{ML-IDS}. Five different machine learning algorithms that use features related to power consumption and network metrics are evaluated on RPL-based networks that use MRHOF and OF0 objective functions  in~\cite{Foley2020}; whereas, 6LoWPAN compression header data (e.g., destination port, context identifier, etc.) are used by six different machine learning algorithms in~\cite{8270652}. In addition, four ensemble learning approaches against seven types of routing attacks are evaluated in~\cite{8777504}. A neural network-based secure system that instruments the source code of the application at compile time is proposed to detect illegitimate accesses to outbound memory in~\cite{10.1145/2990499}. As a graph-based supervised learning approach, the optimum path forest algorithm is used in~\cite{MI-BGSA} to analyze traffic between LLN and the Internet. In~\cite{TL-GP}, the genetic programming-driven transfer learning approach is proposed to develop an IDS that is resistant to change in attack and node types. In addition to machine learning-based approaches, statistical intrusion detection systems are used in~\cite{Shafique2018,Stephen-2018} to detect rank attacks by evaluating the energy states of the nodes. Knowledge-driven IDS is also proposed in~\cite{Midi2017KalisA} to select the optimal set of detection techniques by dynamically collecting knowledge from the network. Here, a wide variety of RPL attacks is considered and a subset of the detection techniques of these attacks is chosen.

Researchers have also proposed IDSs that rely on a distributed placement strategy in which dedicated nodes monitor the network to eliminate the concerns posed by centralized placement. The most satisfactory approach here is to place the IDS on all nodes, which is not applicable for LLNs however, as it brings about a dramatic burden on the network and the nodes. Therefore, the IDS system must consume less energy and storage in this strategy. Additionally, determining the monitoring nodes in this architecture is another matter that should be investigated. Specification-based IDS systems relying on a finite-state machine that is implemented at each monitor node are proposed in~\cite{6098218,info7020025}. A trust-based RPL routing protocol is introduced in~\cite{7878793} for black hole attacks by calculating the trust value for each of the neighboring nodes of the monitoring nodes. An adaptive threshold-based solution is proposed in~\cite{6934253} to ignore erroneous header options, and hence to abolish the inconsistency in the DODAG tree.  A mitigation scheme, named Secure-RPL, that is effective for both static and dynamic network is developed in~\cite{Verma1}. In \cite{Kasinathan2013DenialofServiceDI} they proposed that the system monitors the network traffic of 6LoWPAN through one or more IDS's operating in promiscuous mode. 

Another approach is proposed in~\cite{7930501} where the position of the monitoring nodes is investigated and an RPL-based distributed monitoring strategy is proposed. However, only the version number attack is investigated in~\cite{7930501}, while seven different attacks are targeted in the current study. In addition, they assume that monitoring nodes have higher capacity which leads to a reduction in the load on the other nodes with constrained resources. This makes another major distinction on the communication strategy of monitoring nodes, where they communicate with the root node through a separate RPL instance in~\cite{7930501}. They choose monitoring nodes in which the communication ranges of all the monitoring nodes cover all the other nodes. While they only explore the use of the root node as the central IDS~\cite{7930501}, the location of the central IDS is also explored in this current study. To sum up, different architectures (CIDwL, CIDwG, and DCID) are covered comprehensively with different attacks and attacker locations in the current study.

Hybrid approaches have also been studied to take advantage of both centralized and distributed placement strategies. SVELTE ~\cite{svelte} is the first IDS proposed for LLNs that uses a centralized IDS at the root node and a distributed firewall at every node. An anomaly and specification agent-based IDSs located in, respectively, the root and router or leaf nodes are proposed in~\cite{BOSTANI201752}. Here, the router and leaf nodes in LLN collect information, determine illegitimate nodes, and transmit the analysis results to the root node, which finally evaluates the anomalies in the network. Hybrid architectures in IDSs can potentially achieve higher detection rates while reducing false positives. These architectures provide a balance between detection accuracy and resource efficiency. Centralized components handle the intensive processing tasks, while distributed or cooperative elements ensure scalability and responsiveness. However, hybrid architectures may encounter communication delays and overhead, particularly when aggregating data from a large number of nodes. This challenge becomes more intensified in large networks with limited resources.

Only a few studies analyze the effects of attacks in RPL. Although comprehensive, they generally focus on a particular type of attack, such as version \cite{mayzaud2014study}\cite{aris2016rpl}, rank attacks \cite{le2013impact}. Furthermore, they only analyze the effect of the location of the attackers~\cite{le2013impact,mayzaud2014study,aris2016rpl} or the percentage of the attackers~\cite{dogan2022analysis} on the network, not their detectability by considering their locations relative to the placed IDSs in the network. To the best of the authors’ knowledge, this is the first study to analyze the placement of IDS in a DODAG topology, which is the main contribution of this study.

In addition to traditional approaches, FL methods have also been developed and used as promising solutions considering data sharing and security concerns. A study focuses on detecting DDoS attacks within a fog computing architecture using FL is given in~\cite{li2021fleam}. It highlights that fog/edge computing reduces latency and improves efficiency compared to cloud servers. It also notes that fog/edge computing and federated learning increase the costs for attackers, making attacks more expensive. A few studies~\cite{9146846, man2021intelligent} have focused on the use of convolutional neural networks (CNNs) in industrial IoT (IIoT) to decrease communication burdens, thereby making FL more viable in bandwidth-constrained environments. Another study~\cite{network3010008} employs both CNN and recurrent neural network (RNN) models for centralized machine learning (ML) and FL for accurately identifying unwanted intrusions in wireless IIoT network. An FL-based IDS system for Industry 4.0 scenarios has been proposed in \cite{9120761}. This system employs smart filters at IoT gateways to detect and prevent cyberattacks using deep learning models trained on locally collected data. These models are then shared across gateways to improve detection accuracy, reduce data transmission, and protect privacy. A FL-based IDS systems using differential privacy for IIoT have been proposed in~\cite{chathoth2021federated, 9609643}. 

A Stackelberg game-based methodology designed to promote an incentive mechanism for FL was proposed in~\cite{f7fce76b0f4343e5a816fc41b9f412a9}. This mechanism aims to enhance the interaction between IoT devices and the edge server, facilitating a more efficient and cooperative FL environment. An FL-based strategy specifically for network intrusion detection, aiming to tackle the dual challenges of insufficient network intrusion detection datasets and the need for robust privacy protection, was proposed in~\cite{https://doi.org/10.1002/cpe.6812}. Their approach utilizes the gated recurrent units (GRU) deep learning model for local iterative training on devices. The critical role of distributed learning and particularly the application of FL in expansive IoT networks are discussed in~\cite{9712615}. They noted that while FL offers a promising avenue for integrating ML within large-scale IoT infrastructures, maintaining ML accuracy requires frequent updates to the global algorithms, which cost a significant amount of data.

An FL-based study for detecting wormhole attacks in RPL is presented in \cite{ALGHAMDI2023103014}. Here, the local trainers (representing all nodes) build a global model using long short term memory (LSTM) with eight RPL-related features. Once the model is built, nodes are conditionally activated to locally detect wormhole attacks. This activation depends on the trust value; a node triggers the global model when its cumulative trust value drops below a predefined threshold, thereby extending the system's lifetime. The primary objective of FL in~\cite{ALGHAMDI2023103014} is to maintain data security and privacy among the collaborators.

To sum up, the FL-based studies in IoT generally tend to use publicly available datasets like KDD99~\cite{kdd_cup}, CICIDS2017~\cite{sharafaldin2018toward}, and even MNIST~\cite{mnist}, which either belong to different domains or do not incorporate traffic covering RPL-specific attacks such as decreased version and DAG inconsistency. The only study that addresses attacks in RPL-based network is proposed in~\cite{ALGHAMDI2023103014}. However, they only cover wormhole attacks and their design heavily relies on utilizing all nodes as collaborators. Moreover, their evaluations were conducted in a network environment with only 15 nodes, where they might not even show multihop characteristics~\cite{kim2017challenging}. In this study, we proposed an FL-based IDS for RPL, which cover various attacks against RPL. It has been rigorously evaluated and compared with other ID architectures and the other FL-based studies in the literature. To the best of our knowledge, this is the first study to analyze the placement of IDS in RPL, focusing on a comparative evaluation of FL-based IDS for RPL attacks in a constrained LLN environment.

\section{IDS Architectures}
\label{section_architecture-analysis}
Intrusion detection systems developed for the IoT-based network are mainly categorized from different perspectives (e.g., placement strategy, detection method, response mechanism) in the literature~\cite{zarpelao-taxonomy,Seyfollahi-taxonomy}. Among them, placement is one of the important factors that directly determines the efficacy of IDS. In this study, we take into account three IDS architectures that rely on centralized and distributed placement strategies, and we evaluate their advantages and disadvantages when they are subject to different routing attacks. These IDS architectures are briefly explained in the following sections, and evaluations regarding their performance are given in detail later.

\subsection{Central ID with Local Information (CIDwL)}
\label{section_architecture-analysis-standalone}
Here, the IDS is placed in a centralized manner, where a border router or a dedicated node is responsible for monitoring traffic and raising alarm. The main advantage of this architecture is that it does not bring about additional communication costs. Moreover, the lifetime of the network is not affected much since only a single node, in most cases a powerful device, is responsible for monitoring the traffic and extracting features. The downside, however, is the single point of failure phenomena that occurs when the IDS node is down by an intruder, leading to a network that is open to harmful attacks. In addition, the IDS node reaches a decision based on its local data naturally collected. 

\subsection{Central ID with Global Information (CIDwG)}
\label{section_architecture-analysis-cooperative}
In this architecture, the IDS node is also placed in a central node. However, other nodes might participate in intrusion detection by forwarding some features extracted from their local traffic. Hence, the central IDS node, who makes the decision, has richer information about the network. However, the single point of failure is still a major handicap of this approach. In addition, this approach results in an increase in communication cost as the extracted local features are periodically sent to a central ID node and reduces the average lifetime of the network.

\subsection{Distributed and Collaborative ID (DCID)}
\label{section_architecture-analysis-distributed}
Unlike previous architectures, more than a single ID node is dedicated so that they function independently. Therefore, each ID node individually performs \textit{i}) traffic monitoring, \textit{ii}) feature extraction, and \textit{iii}) making decisions and raising local alarms. Local alarms are transmitted by ID nodes to the root node, where they are aggregated. It is worth noting that this aggregation can also take place at the edge or in the cloud. Then, the root node could trigger a global alarm based on a threshold value of local alarms.

\section{Comparative Analysis of Architectures}
\label{section_comparative-analysis-architecture}
There are a number of factors that directly affect the performance of IDS developed for IoT networks. These include detection model, adopted architecture, attack types, attacker locations, and the like. Among them, no doubt, the IDS architecture plays a key role as it enables the ID model to function both effectively and efficiently. Therefore, the evaluation of the optimal IDS architecture is one of the important research directions today. Here, we explore the role of three architectures from the point of view of intrusion detection capability, communication cost, and security.

\subsection{Simulation Setting and Environment}
\label{section_architecture-analysis-simulation-setting}
\begin{figure}[!h]
    \centering
    \includegraphics[width=\linewidth]{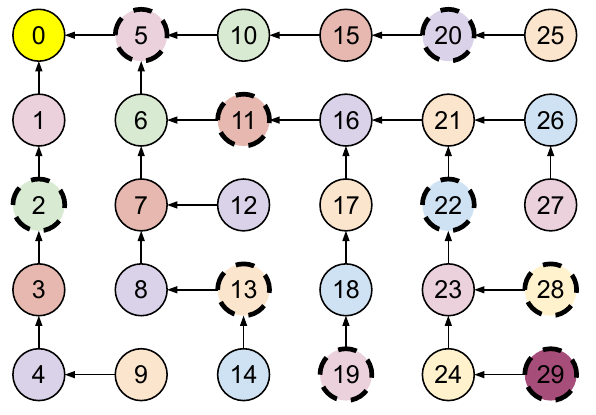}
    \caption{The network topology}
    \label{fig:topology}
\end{figure}

The grid topology, shown in Figure~\ref{fig:topology}, is used in the experiments. As seen here, we have deployed 30 nodes, including the root node in the simulations because at least 25 nodes are suggested in RPL-based networks to see the multihop characteristics of RPL~\cite{kim2017challenging}. 

\begin{figure}[!h]
    \centering
    \includegraphics[width=\linewidth]{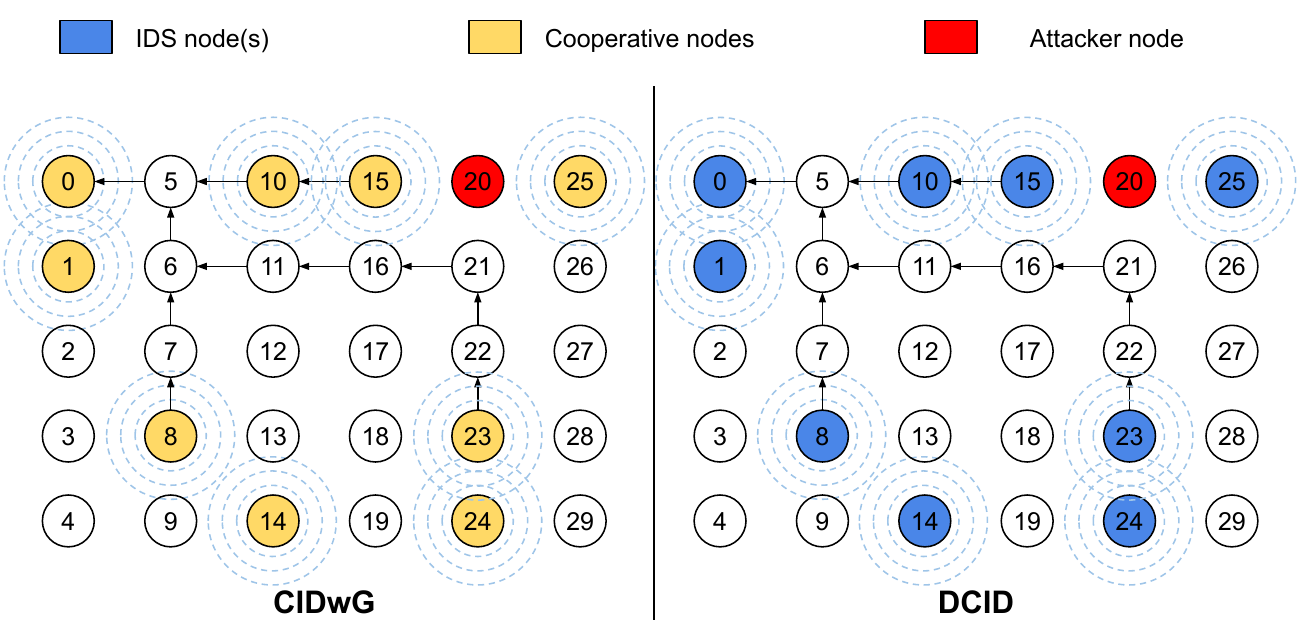}
    \caption{A scenario with nine collaborators for CIDwG and DCID architectures}
    \label{fig:exemplar_scenario}
\end{figure}

The nodes in the DODAG shown in the figure are leveled according to their ranks, and each level is represented with different face colors in the figure. As seen, there are 10 levels to involve 30 nodes in the DODAG. The nodes are 20 m away from each other, and the transmission range in the network is set to 25 m so that each node can communicate only its one-hop neighborhood in the DODAG. The arrows in Figure~\ref{fig:topology} represent the preferred parent of child nodes in DODAG under benign environment. The dashed borders in the figure, however, represents the attacker nodes. It is worth noting here that attacker nodes are arbitrarily chosen from every level in DODAG. Taking this DODAG into account, several network scenarios are generated separately for the three types of architecture. An exemplar scenario for the CIDwG and DCID architectures is separately given in Figure~\ref{fig:exemplar_scenario}. As illustrated, up to nine nodes from each DODAG level are randomly selected to act as collaborators. In both architectures, these nodes extract features, either forwarding them to the root node (in CIDwG) or using them to generate local intrusion detection systems (in DCID). Further details on these architectures are provided in the next section.
    
In order to set up the architectures and implement routing attacks, the Cooja simulator~\cite{COOJA}, a Java-based simulator that emulates sensor notes running the Contiki operating system~\cite{contiki-ng} (version 2.7), is used in the experiments. Cooja supports simulating wireless sensor networks involving different mote types, and here we adopted the Zolertia Z1 platform due to its larger memory capacity than the other platforms. This enables us to implement attacks on Contiki OS. The parameter values that we adopt in the simulations are listed in Table~\ref{table_simulation_parameters}.

\begin{table}[!h]
	\centering
	\caption{Simulation Parameters}
	\begin{tabular}{ll}
		\toprule
		\textbf{Parameters}&\textbf{Values} \\ \midrule
		   Radio Environment & UDGM: Distance Loss\\
		Objective Function & MRHOF-ETX\\
		TX Range & 25 m \\
		Simulation Time & 5 hours \\
		Area of Deployment & 100 m $\times$ 80 m \\
		Number of Sink Node	& 1 node \\
		Number of Sensor Node & 29 nodes\\
		Traffic Pattern & UDP packets sent every 15 s \\
		\bottomrule
	\end{tabular}
	\label{table_simulation_parameters}
\end{table} 

In the experiments, we employ a machine learning-based detection method in order to classify network traffic as `malicious' or `benign'. Here, the random forest (RF) algorithm~\cite{randomforest}, one of the most popular ensemble learning algorithms, is used for each architecture and scenario. We have used `scikit-learn'~\cite{scikit-learn} library for the implementation of the RF algorithm. The default parameters of the RF algorithm adopted in the library are used in the experiments, and the 10-fold cross-validation is applied. As for the input to train the RF algorithm; the entire traffic flow is first split into 60 s time windows by the nodes in charge, and the features are then extracted from each windowed interval. Note that we have adopted 35 features that are proposed in~\cite{TL-GP}. These features are extracted from the network layer, primarily incorporating the count and frequency of data and routing control packets as well as RPL-related features such as version and rank numbers within these control packets. Finally, the generated models are evaluated by the following metrics: accuracy, true positive rate (TPR), and false positive rate (FPR).

\subsection{Evaluations of Architectures}
\label{section_architecture-research-questions}
We aim to evaluate three IDS architectures against seven routing attacks by seeking to answer the following three research questions (RQs). The \textit{i}) main motivation, \textit{ii}) the method adopted in the experiment, and finally \textit{iii}) the findings are successively explained in detail for each RQ below. Note that the experiments are conducted on the basis of these RQs. In order for the experiments to be replicable by the researchers, the data used throughout the experiment are shared\footnote{https://wise.cs.hacettepe.edu.tr/projects/IDArch/results.zip}. 
    
\noindent\textbf{RQ1: Is one central IDS enough for effectively detecting all types of attacks that are performed at different locations?}

\textbf{Motivation:} RPL attacks have a varying effect on network traffic and nodes. Depending on their characteristics as well as the position of the attacker node with respect to the root node and the ID node, these attacks can harm a very limited scale or, if not the entire, a very large portion of the network. Therefore, prior to the development of a detection system in a central IDS architecture, it is certainly important to study the optimal or at least near-optimal position of the ID node relative to the location of the attacker node in the network for the types of attacks that the system may be subject to. Here, we evaluate the position of the ID node as a function of the distance from the root node and the attacker node for each routing attack. We believe that this attempt will give an important insight to network security practitioners on the optimal placement of IDS when developing a security solution in the central architecture.  
   
\textbf{Method:} In order to reach this goal, we have conducted a massive experiment in which a great number of network scenarios are generated based on the CIDwL architecture. Here, we adopt a strategy to enroll a single node as the ID node and another node as the attacker node. As stated earlier, we used the grid topology shown in Figure~\ref{fig:topology} in the experiments. From this topology, we select a node from each DODAG level and set them to the attacker node. Therefore, given in order of DODAG level, the nodes 5, 2, 11, 20, 13, 22, 19, 28, and 29 are taken as an alternative to attacker node (they are represented by the dashed line in the figure). As for the ID nodes, there are 30 nodes (from node 0 to node 29) that can be an alternative to the ID node. For each attack type, we take the combination of all the alternatives of the node selection cases to generate the scenarios in the experiment; therefore, 1,827 network scenarios are generated in total. It is worth stating here that the occurrences in the combination where the IDS and attacker nodes point to the same node are excluded.

\begin{figure*}[h]
\includegraphics[width=\textwidth]{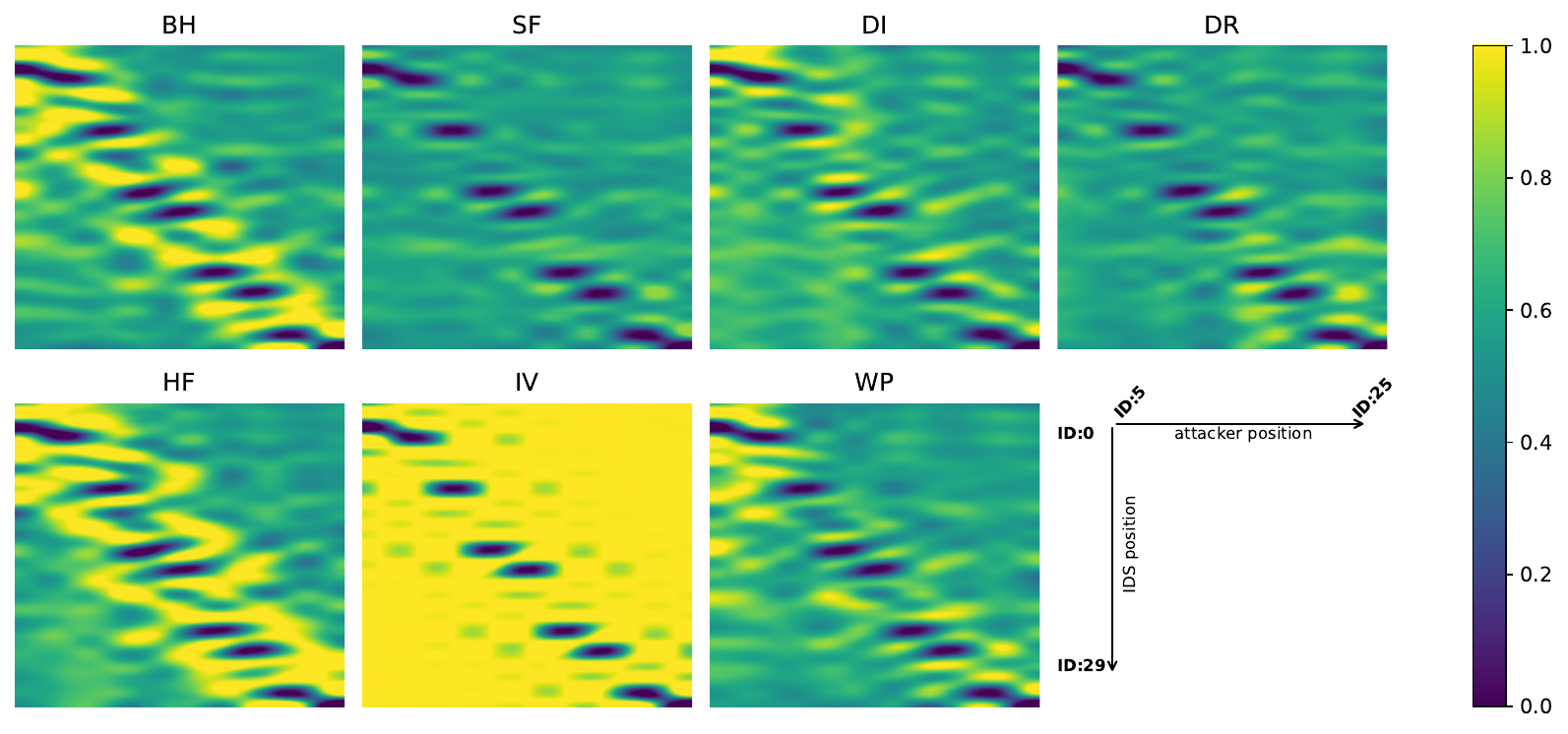}
\caption{The detection accuracy performance of IDSs with respect to the attacker location in CIDwL architecture. The x- and y-axis represent, respectively, the location of attacker and IDSs nodes (given in order of distance from root to leaf).}
\label{fig:ids_performance_standalone}
\end{figure*}

\textbf{Results:} The detection accuracy is taken into account to evaluate the performance of IDS under the CIGwL architecture. These detection performances are individually shown for each attack with a heat map given in Figure~\ref{fig:ids_performance_standalone}. In the figure, the x- and y-axes represent the positions of the attacker and the ID nodes in the topology, respectively. Therefore, the upper-left and lower-right parts of the maps simply represent the regions in which both the attacker and the ID nodes are closest and farthest to the root node, respectively; the major diagonal part, however, represents that the ID node is only a few hops away from the attacker node (or vice versa). Beware that the detection performances of the scenarios where the position of ID node and attacker represents the same node (e.g., node 5) are evaluated as `\textit{0.0 accuracy}' in the heat maps because they are not studied. So, they can be ignored in the figure.
    
As it is clearly seen in Figure \ref{fig:ids_performance_standalone} that the performance of IDS on CIDwL architecture is highly dependent on the attack type, the distance to the root node, and also the distance to the attacker node. As for the evaluation on an attack basis; no matter wherever the ID or attacker nodes are positioned, the detection accuracy of IDS for the SF attack is very poor (max 74\% accuracy observed); while it is remarkably high for the IV attack (min 99\% accuracy observed). This proves that the ID model fails to discriminate malicious traffic from benign traffic by examining the feature set for SF attack. It is by no means surprising because the data packets may already be dropped in a typical wireless medium, hence features from lower layers might be needed for effective detection of SF~\cite{canbalaban2020cross}. On the contrary, the ID node can easily detect IV attack as the intruder globally alters the version number in DIO packet which makes it an easy-to-detect attack in this study, particularly with the version-related features (e.g., \texttt{MAX\_VERSION}). This also emphasizes the contribution of our study compared to other studies analyzing only one attack (version attack) \cite{7930501}.
    
The locations of nodes are important when evaluating for other types of attacks (i.e.; BH, HF, DI, WP, and DR). Actually, the detection performance increases dramatically when the ID node is positioned a few hops away from the attacker node for BH and HF attacks. Because the BH attack leads to a continuous drop of data packets sent by the child node, and the HF attack emerges overwhelming DIS packets towards the nodes that are neighbors of the attacker, they mainly harm the attacker's locality in the network. That's why, as the findings suggest, the ID nodes are better to be positioned near the attacker nodes for these two types of attacks. As for DI and WP attacks, IDS also shows a satisfactory detection performance when it is close not only to the attacker node but also to the root node. This is due to the nature of these attacks. Speaking concretely, the WP attacker intentionally selects its worst parent node, leading to data packets being forwarded through a suboptimal route. Because data traffic is concentrated at the root node, the WP attacker near the root node yields even a much higher deviation in terms of misrouted data traffic compared to the benign environment. Therefore, the ID node in the attacker's locality can easily capture malignant traffic by analyzing the change in the number of data packets it forwards in malicious and benign environments. This is very alike to what happens with the DI attack where, due to the illegitimate setting of the flags, almost all data packets are dropped, especially when the attackers are near the root node. In contrast, the DR attack can be only detected effectively when the ID and attacker nodes are far away from the root node but close to each other. This is because of the objective of the attack, that is, misleading the benign nodes by illegitimately advertising them a lower rank value. The nodes near the root node have already lower rank values in DODAG, and the attackers are unable to generate a deviation compared to the benign environment when they are also near the root node. In such a circumstance, the ID node fails to discriminate whether the change in rank values is legitimate or not.
    
In addition to a general discussion of the CIDwL architecture, we also assess to which degree the attacker position can affect the detection accuracy of IDS under the central architecture. To do that, we consider the scenarios where the IDS is placed to the root node (node 0) because the plethora of the studies in the literature adopts the placement of IDS to the root node so that root node can analyze all the traffic exchanged between LLN and the Internet~\cite{zarpelao-taxonomy}. The accuracy performances reported separately for each attacker location are shown in Table~\ref{tbl:root_node_performance}. Note that, the best-performing accuracy values are highlighted with the gray color. The last column indicates the major difference in the accuracy performance. The results reveal that the degradation on the accuracy performance of IDS reaches from 13.17\% (DR attack) to 49.50\% (DI attack) for attacks where the optimal positioning of IDS is important (i.e., BH, DI, DR, HF, and WP). Therefore, it can be concluded that satisfactory detection performance with a single central ID can only be obtained when the position of the attacker is close to the IDS, maximum 2-levels away as shown in Table \ref{tbl:root_node_performance}. In addition to the accuracy performance, Table~\ref{tbl:tpr_fpr_standalone} gives the best and worst TRP and FPR performances corresponding to the best and worst accuracy values in Table~\ref{tbl:root_node_performance}, respectively, to also reveal the performance difference ($\Delta$) in terms of detection rate and false alarm rate. The results are highly correlated with those of Table~\ref{tbl:root_node_performance}, and the difference between TPR and FPR can reach 51\% and 48.7\%, respectively.

\begin{table*}[!h]
    \centering
    \caption{Detection accuracy of the root node (ID:0) when attackers are at different levels (L)}
    \begin{tabular}{ccccccccccc} \toprule
    \multirow{2}{*}{\textbf{Attack}} & \multicolumn{9}{c}{\textbf{Attacker IDs}}                                                                                 & \multirow{2}{*}{\textbf{Major Difference}} \\ \cline{2-10}
                                     & \textbf{5 (L1)} & \textbf{2 (L2)} & \textbf{11 (L3)} & \textbf{20 (L4)} & \textbf{13 (L5)} & \textbf{22 (L6)} & \textbf{19 (L7)} & \textbf{28 (L8)} & \textbf{29 (L9)} &                                            \\ \midrule
    \textbf{BH}                      & \cellcolor{gray25}0.988      & 0.832      & 0.533       & 0.575       & 0.530       & 0.535       & 0.540       & 0.515       & 0.562       & 0.473                                      \\
    \textbf{SF}                      & \cellcolor{gray25}0.580      & 0.507      & 0.528       & 0.577       & 0.540       & 0.505       & 0.508       & 0.540       & 0.512       & 0.075                                      \\
    \textbf{DI}                      & 0.967      & \cellcolor{gray25}0.992      & 0.557       & 0.687       & 0.520       & 0.553       & 0.570       & 0.497       & 0.503       & 0.495                                      \\
    \textbf{DR}                      & 0.500      & 0.515      & 0.568       & 0.568       & 0.535       & \cellcolor{gray25}0.632       & 0.528       & 0.593       & 0.538       & 0.132                                      \\
    \textbf{HF}                      & \cellcolor{gray25}1.000      & \cellcolor{gray25}1.000      & 0.802       & 0.667       & 0.582       & 0.560       & 0.592       & 0.517       & 0.553       & 0.483                                      \\
    \textbf{IV}                      & 0.998      & \cellcolor{gray25}1.000      & \cellcolor{gray25}1.000       & 0.998       & 0.998       & 0.998       & 0.997       & 0.997       & 0.993       & 0.007                                      \\
    \textbf{WP}                      & \cellcolor{gray25}0.927      & 0.805      & 0.517       & 0.503       & 0.510       & 0.575       & 0.522       & 0.537       & 0.493       & 0.433         \\ \bottomrule                     
    \end{tabular}

    \label{tbl:root_node_performance}
\end{table*}

\begin{table}[!h]
\caption{Best and worst TPR and FPR performances of the root node}
\begin{tabular}{llllllll} \toprule
\multirow{2}{*}{\textbf{Attack}} & \multicolumn{2}{c}{\textbf{TPR}} & \multirow{2}{*}{\textbf{$\Delta_{TPR}$}} & \textbf{} & \multicolumn{2}{c}{\textbf{FPR}} & \multirow{2}{*}{\textbf{$\Delta_{FPR}$}} \\ \cline{2-3} \cline{6-7}
                                 & \textbf{best}  & \textbf{worst}  &                                & \textbf{} & \textbf{best}  & \textbf{worst}  &                                \\ 
                                 \midrule
\textbf{BH}                      & 0.983          & 0.523           & 0.460                          &           & 0.007          & 0.493           & 0.487                          \\
\textbf{SF}                      & 0.553          & 0.493           & 0.060                          &           & 0.393          & 0.483           & 0.090                          \\
\textbf{DI}                      & 0.997          & 0.487           & 0.510                          &           & 0.013          & 0.493           & 0.480                          \\
\textbf{DR}                      & 0.653          & 0.460           & 0.193                          &           & 0.390          & 0.460           & 0.070                          \\
\textbf{HF}                      & 1.000          & 0.500           & 0.500                          &           & 0.000          & 0.467           & 0.467                          \\
\textbf{IV}                      & 1.000          & 0.990           & 0.010                          &           & 0.000          & 0.003           & 0.003                          \\
\textbf{WP}                      & 0.943          & 0.467           & 0.477                          &           & 0.090          & 0.480           & 0.390   \\ \bottomrule                      
\end{tabular}
\label{tbl:tpr_fpr_standalone}
\end{table}
    
\noindent\textbf{RQ2: What is the minimum number of IDS required for effective detection?}
    
\textbf{Motivation:} In the CIDwL architecture, only one node is running as the second defence of line. While this brings about some advantages such as lower communication cost, increased network lifetime, the performance of IDS depends on the type of attack and the attacker position, which is clearly shown by the findings in RQ1. Hence, this architecture is not scalable, and a distributed architecture seems more suitable for RPL-based IoT in order to obtain higher detection accuracy. For this purpose, here we consider CIDwG architecture where more than one node take part in intrusion detection. Unlike the CIDwL architecture, running this architecture increases \textit{i}) computational cost (and consequently the average power consumption) and \textit{ii}) communication overhead, since participating ID nodes are responsible for monitoring and extracting the features and sending to a central ID node. Here, we mainly attempt to reveal how much the detection performance improves by including other nodes in intrusion detection. We also try to answer how many nodes are enough for effective detection and explore trade-offs between detection accuracy and cost (i.e., power consumption and communication cost). 

\textbf{Method:} The main task is to generate scenarios that rely on CIDwG architecture using the network topology shown in Figure~\ref{fig:topology}. Because it is not applicable to make all the nodes as participating nodes for a realistic scenario, we have chosen one node from each DODAG level as ID (except attackers as in RQ1). Therefore, the nodes 0, 1, 10, 15, 8, 25, 14, 23, and 24 (given in order of DODAG level) are selected randomly as the participating nodes in CIDwG. These nodes are demonstrated in Figure~\ref{fig:exemplar_scenario}. The combination of all these nodes (with lengths of two to nine) is generated to obtain participating ID node sets (i.e., \{0, 1\}, \{0, 10\}, \dots, \{0, 1, 10\}, \dots, \{0, 1, 10, 15, 8, 25, 14, 23, 24\}). By adopting this strategy, we have generated 31,626 scenarios in total for CIDwG architecture.  
  
\textbf{Results:} Firstly, the overall accuracy performance of CIDwL and CIDwG as a function of the number of ID nodes are compared. To do that, we first obtained the accuracy performances of CIDwG architecture separately for different attacker locations, and then grouped them with respect to the number of ID nodes in the network. Then, we take the average of each of these groups to reveal the overall accuracy performance of the CIDwG architecture based on the number of participating nodes. The same procedure is also applied for CIDwL architecture except for the grouping as only one ID node is enrolled there. Please note that, in order to ensure a fair comparison, for the CIDwL architecture, only the accuracy performances of the ID nodes used in the CIDwG architecture (i.e., 0, 1, 10, 15, 8, 25, 14, 23, 24) are taken into account.  
    
\begin{figure}[!h]
    \centering
    \includegraphics[width=\linewidth]{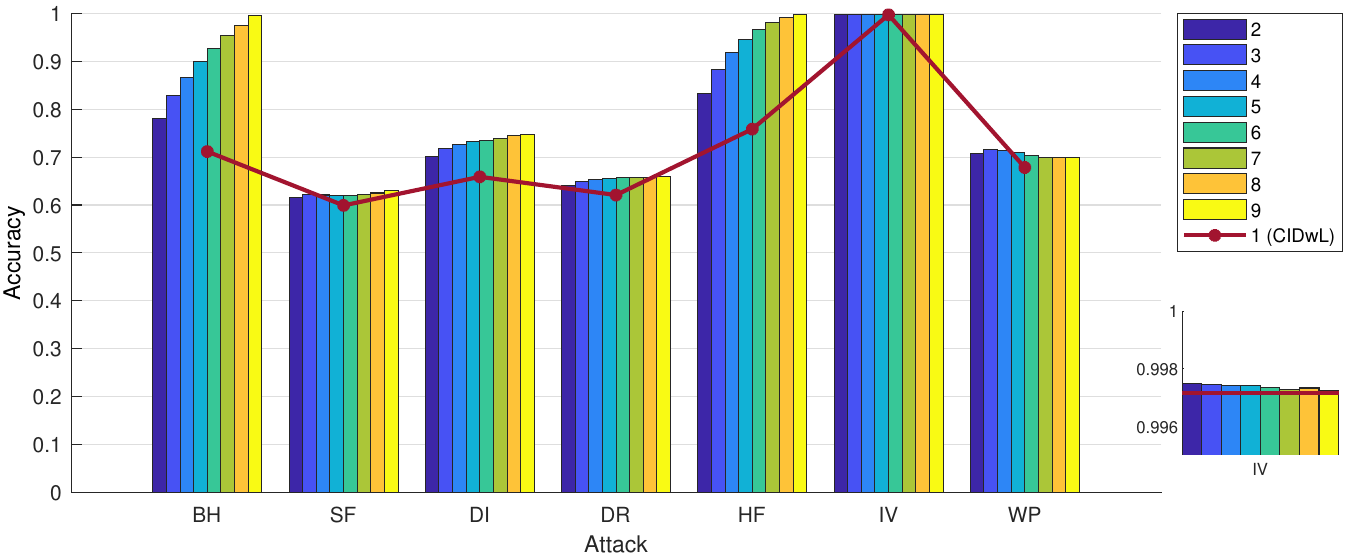}
    \caption{The comparison of the overall accuracy obtained by CIDwL and CIDwG architectures}
    \label{fig:cooperative_accr_number_ids}
\end{figure}

The overall performance of the CIDwL and CIDwG architectures is shown comparatively in Figure~\ref{fig:cooperative_accr_number_ids}. Note that the line and bar plot shows the overall accuracy performances of CIDwL and CIDwG architectures, respectively. Therefore, it gives not only the comparative overall performances of the architectures, but also the change in the accuracy according to the number of ID nodes in CIDwG architecture. As seen here, the performance improves as the number of nodes increases for BH, SF, DI, DR, and HF attacks in CIDwG architecture. For the IV attack, however, satisfactory performance is already achieved with only two ID nodes, with a minimal reduction of $2.7\times10^{-4}$, highlighted in the magnified inset. As for the comparison of the architectures, CIDwG architecture yields better performance than the CIDwL architecture for all types of attacks, where the accuracy difference of the best performances ranges from less than 1\% (for IV attack) to 29.33\% (for BH attack). 

The TPR and FPR performances of the CIDwG architecture with two and nine participating nodes that give the worst and the best accuracy performance, respectively are also analyzed in detail. These performances are given in Table~\ref{tbl:tpr_fpr_CIDwG}. It is also seen here that attacks have a strong impact on the performance of TPR and FPR, and the difference in the performance of TPR and FPR reaches 21.0\% (for BH attack) and 21.8\% (for BH attack), respectively when the IDS is developed in CIDwG architecture with nine participating nodes. The highest variations in BH attacks are primarily due to the nature of the attack. Since the attacker drops data packets instead of forwarding them, it significantly reduces the number of packets that should reach the root node. As a result, the change in forwarded packets at different levels of the DODAG structure becomes a key indicator of the attack. Involving more collaborators at each hop of the DODAG enables the root node to better detect these changes.  Conversely, fewer collaborators, especially those near the root, may fail to detect packet forwarding changes, as packets could have already been dropped at higher hops, or the attackers might be positioned closer to the root. Similarly, the HF attack shows variations in TPR and FPR as the number of collaborators increases, due to a local burst of DIS packets followed by DIO packets. Adding more collaborators helps the root node detect the attack by sharing changes in control packet numbers. 

While the participation of additional nodes improves detection (e.g., nine nodes are needed to effectively detect the BH attack), it also introduces additional communication and computation costs to the network. Therefore, the trade-off between accuracy, TPR, FPR, and network performance must be carefully balanced, as discussed in the next research question.
    
\begin{table}[!h]
\centering
\caption{Comparative TPR and FPR values obtained with two and nine participating nodes in the CIDwG architecture}
\begin{tabular}{llllll} \toprule
\multirow{2}{*}{\textbf{Attack}} & \multicolumn{2}{c}{\textbf{with two nodes}}                         & \textbf{} & \multicolumn{2}{c}{\textbf{with nine nodes}}                        \\ \cline{2-3} \cline{5-6}
                                 & \multicolumn{1}{c}{\textbf{TPR}} & \multicolumn{1}{c}{\textbf{FPR}} & \textbf{} & \multicolumn{1}{c}{\textbf{TPR}} & \multicolumn{1}{c}{\textbf{FPR}} \\ \midrule
\textbf{BH}                      & 0.783                            & 0.222                            &           & \cellcolor{gray25}0.994                            & \cellcolor{gray25}0.003                            \\
\textbf{SF}                      & 0.605                            & 0.375                            &           & \cellcolor{gray25}0.618                            & \cellcolor{gray25}0.358                            \\
\textbf{DI}                      & 0.694                            & 0.292                            &           & \cellcolor{gray25}0.744                            & \cellcolor{gray25}0.249                            \\
\textbf{DR}                      & 0.630                            & 0.347                            &           & \cellcolor{gray25}0.640                            & \cellcolor{gray25}0.320                            \\
\textbf{HF}                      & 0.831                            & 0.165                            &           & \cellcolor{gray25}1.000                            & \cellcolor{gray25}0.003                            \\
\textbf{IV}                      & 0.997                            & \cellcolor{gray25}0.002                            &           & \cellcolor{gray25}0.998                            & 0.003                            \\
\textbf{WP}                      & \cellcolor{gray25}0.706                            & \cellcolor{gray25}0.292                            &           & 0.704                            & 0.304    \\ \bottomrule                  
\end{tabular}
\label{tbl:tpr_fpr_CIDwG}
\end{table}
    
Here, the best performance of each architecture is also given. To do that, separately for each attack type, we first report the maximum accuracy and its corresponding number of participating nodes obtained with respect to the attacker location. Then, the average of the best detection performances for all different attack positions in each attack type is taken for comparison of CIDwL and CIDwG. Again, for the CIDwL architecture, only the accuracy performances of the ID nodes used in the CIDwG architecture (i.e., 0, 1, 10, 15, 8, 25, 14, 23, 24) are taken into account.  
    
The average of the best accuracy performances is shown in Table~\ref{tbl:RQ2_results}. The detailed best performances on the attacker location basis are also reported separately for each attack type in\footnotemark[\value{footnote}]. Note that the best-performing architecture is represented with gray shades in the table. It can easily be seen from the table that very similar detection performance is observed for BH, HF, and IV attacks, while superior detection performance is observed when IDS is developed with the CIDwG architecture for SF, DI, DR, and WP attacks. The overall improvement on the detection accuracy with the CIDwG architecture ranges from less than 1\% to around of 2.7\% on overall. While the improvement is not dramatic, please note that the best performances are obtained with different placements of ID nodes each time. Hence, it is not applicable in real life. The best results also show that, depending on the attack type and attacker location, two to five ID nodes are enough to reach the best performances with CIDwG architecture. 
    
\begin{table}[!h]
\centering
\caption{Average of the best detection accuracy performances for CIDwL and CIDwG}
\begin{tabular}{lcccccc} \toprule
\multirow{2}{*}{\textbf{Attack}} & \multicolumn{2}{c}{\textbf{Accuracy}} & \textbf{} & \multicolumn{3}{c}{\textbf{Number of Nodes}} \\ \cline{2-3}\cline{5-7}
                                 & \textbf{CIDwL}    & \textbf{CIDwG}   & \textbf{} & \textbf{max}  & \textbf{min} & \textbf{median} \\ \midrule
\textbf{BH}                      & \cellcolor{gray25}0.996                  &  \cellcolor{gray25}0.996                  &           & 3             & 2            & 2               \\
\textbf{SF}                      & 0.697                  & \cellcolor{gray25}0.715                  &           & 4             & 2            & 3               \\
\textbf{DI}                      & 0.847                  & \cellcolor{gray25}0.856                  &           & 5             & 2            & 2               \\
\textbf{DR}                      & 0.718                  & \cellcolor{gray25}0.744                  &           & 5             & 2            & 4               \\
\textbf{HF}                      & \cellcolor{gray25}0.999                  & \cellcolor{gray25}0.999                  &           & 2             & 2            & 2               \\
\textbf{IV}                      & 0.999                  & \cellcolor{gray25}1.000                  &           & 3             & 2            & 2               \\
\textbf{WP}                      & 0.845                  & \cellcolor{gray25}0.852                  &           & 5            & 2            & 2           \\ \bottomrule
\end{tabular}
\label{tbl:RQ2_results}
\end{table}

\noindent\textbf{RQ3: How do IDSs make decision together efficiently from the communication cost perspective?}
    
\textbf{Motivation:} We show that enrolling multiple nodes as part of IDS dramatically increases the detection accuracy on overall. However, as stated earlier, involving additional nodes get down the network's performance from the point of communication cost. This is because LLNs are characterized by low data rates, limited frame size, and high packet loss~\cite{rfc6550}. When considering a scenario in which a number of participating nodes periodically transmit the local features through a sequence of fragmented frames, the effectiveness of the CIDwG architecture is highly questionable. Moreover, because these nodes have to transmit their local information within a much shorter interval to ensure `early' detection capability of IDS, the applicability of this architecture on such a network should even be discussed. Therefore, in DCID architecture, the enrolled ID nodes are operating individually and, rather than the massive feature data, only alarms are transmitted periodically. The main goal here is to analyze if DCID architecture yields better than or at least similar detection performance to the CIDwG architecture does. Hence, the aforementioned communication cost, network lifetime problems accelerated by CIDwG architecture could be overcome. The major question to be addressed in DCID architecture is to find optimal voting ratio to raise a global alarm in the network.  
    
\textbf{Method:} As in RQ2, we here have created a large number of scenarios to implement DCID architecture on the network topology given in Figure~\ref{fig:topology}. The same ID nodes as in RQ2 are considered for a fair comparison; and therefore, the combination of these nodes (with a length of two to nine) is generated to obtain distributed node sets. Contrary to the CIDwG architecture, a voting scheme has to be tuned in DCID architecture. It is simply evaluated that of all local alarms arising from the distributed ID nodes, how many are necessary to raise a global alarm. Here, DCID architecture is evaluated with two voting schemes for each attack type: \textit{i}) minority voting and \textit{ii}) majority voting. In minority voting, a global alarm is issued, provided that at least a local alarm is sent by one of the ID nodes. In the majority voting, however, the ratio of local alarms ($R\ | R\in \{50, 60, 70, 80\}$) matters, and a global alarm is issued as long as $R\%$ of the ID nodes arises local alarms. It is worth stating here that because the lower or greater $R$ values (i.e., $R < 50$ or $R > 80$) lead to a performance degradation, we here disregard studying them further. As in RQ2, we have generated 31,626 scenarios for each voting scheme in the DCID architecture.

\textbf{Results:} Because we mainly aim to investigate if the DCID architecture can be a good alternative to the CIDwG architecture, here we compare them with each other in terms of overall accuracy performance. In order to accomplish that, we first find out the best voting scheme that is to be adopted for DCID architecture. To do that, we collect the accuracy results obtained by different voting schemes for different attacks and then calculate the average of these accuracy values. The overall comparative performance of the voting schemes in the DCID architecture is outlined in Table~\ref{tbl:overall_performance_voting_schemes_DCID}, and again the best performances are highlighted in gray. From the comparative results, it is clear that minority and 80\% voting schemes have shown the poorest accuracy, whereas 50\% voting scheme (followed by 60\% voting scheme) has shown the highest accuracy. Therefore, we adopted the 50\% voting scheme for the DCID architecture to compare its overall accuracy performance with that of the CIDwG architecture.

    \begin{table}[!h]
    \centering
    \caption{Comparison of overall accuracy performances of voting schemes in DCID architecture}
    \begin{tabular}{llllll}\toprule
    \multicolumn{1}{c}{\multirow{2}{*}{\textbf{Attack}}} & \multicolumn{5}{c}{\textbf{DCID}}                                                                                                                     \\ \cline{2-6}
    \multicolumn{1}{c}{}                & \multicolumn{1}{c}{\textbf{minority}} & \multicolumn{1}{c}{\textbf{50\%}} & \multicolumn{1}{c}{\textbf{60\%}} & \multicolumn{1}{c}{\textbf{70\%}} & \multicolumn{1}{c}{\textbf{80\%}} \\ \midrule

    \textbf{BH}                                          & 0.614 & \cellcolor{gray25}0.805 & 0.798 & 0.734 & 0.681 \\
    \textbf{SF}                                          & 0.548 & \cellcolor{gray25}0.648 & 0.642 & 0.604 & 0.577 \\    
    \textbf{DI}                                          & 0.592 & \cellcolor{gray25}0.737 & 0.727 & 0.677 & 0.638 \\
    \textbf{DR}                                          & 0.560 & \cellcolor{gray25}0.677 & 0.671 & 0.630 & 0.599 \\
    \textbf{HF}                                          & 0.653 & \cellcolor{gray25}0.866 & 0.857 & 0.787 & 0.728 \\
    \textbf{IV}                                          & 0.996 & \cellcolor{gray25}0.998 & \cellcolor{gray25}0.998 & \cellcolor{gray25}0.998 & 0.997 \\
    \textbf{WP}                                          & 0.597 & \cellcolor{gray25}0.754 & 0.748 & 0.695 & 0.652 \\ \bottomrule
    \end{tabular}
    \label{tbl:overall_performance_voting_schemes_DCID}
    \end{table}

\begin{figure}[!h]
    \centering
    \includegraphics[width=\linewidth]{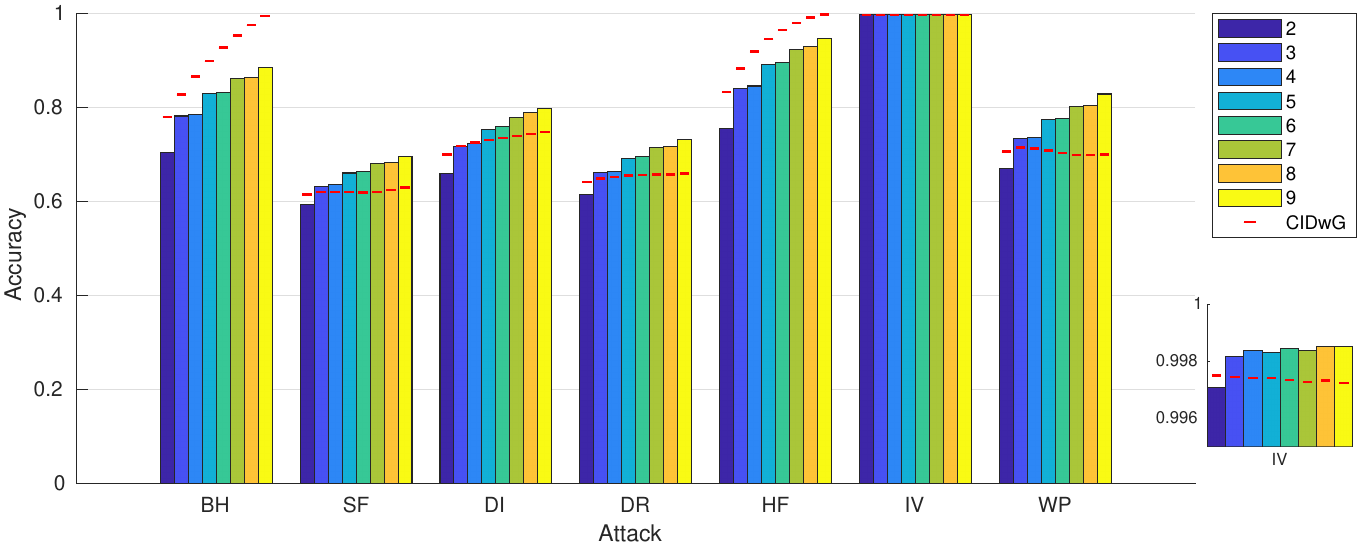}
    \caption{The comparison of the overall accuracy as a function of varying number of nodes in DCID architecture (50\% voting scheme)}
    \label{fig:distributed_accr_number_ids}
    \end{figure}

In order to compare the CIDwG and DCID architectures (with 50\% voting scheme) as a function of the number of distributed nodes, we first collect the accuracy performances obtained with two to nine distributed nodes and then calculate the average of these performances. The overall accuracy performance of the DCID architecture is shown in Figure~\ref{fig:distributed_accr_number_ids}. Note that the results for other schemes can also be found in\footnotemark[\value{footnote}]. It is worth stating that the corresponding accuracy values obtained by the CIDwG architecture (reported in Figure~\ref{fig:cooperative_accr_number_ids}) are also indicated in this figure by horizontal red markers for the sake of a clear performance comparison of two architectures.
   
From the performance with DCID architecture, it can be seen that a better detection accuracy is reached as more distributed nodes are participating for all attack types except IV attack, where a satisfactory performance is already obtained with even two IDs (the highest change is $1.4\times 10^{-3}$). It is important to note that the participation of more nodes also increases the reliability of the DCID architecture. This is because even if the intruder targets and prohibits some of these nodes from operating, the IDS can still function with a detection performance that may degrade to some extent. On the contrary, the major downside of this setting is the reduced average lifetime of the network, as resource-constrained devices periodically extract features. Therefore, a trade-off exists for the DCID architecture that should be properly balanced taking into account the detection performance, reliability of the IDS, and constrained nature of the devices. 
   
When it comes to the comparison of the performances of the architectures, the findings here suggest that, no matter how many nodes are participating, CIDwG outperforms DCID for BH and HF attacks, and a slight performance improvement is also observed when CIDwG architecture with two participating nodes; whereas DCID shows better detection performance for SF, DI, DR, and WP attacks with three or more participating nodes.

The TPR and FPR performances of the DCID architecture with two and nine participating nodes are also evaluated to reveal the change in the performances with these two settings. The results are comparatively given in Table~\ref{tbl:tpr_fpr_dcid} where the best performances are again highlighted in gray. The results here suggest that DCID architecture with two participating nodes yields better performance than that with nine participating nodes in terms of TPR for all types of attack except HF. The difference here ranges from less than 1\% (for the IV attack) to 14.7\% (for the SF attack). Regarding FPR performance, a dramatic improvement is observed for all types of attack when nine nodes participate. The difference here ranges from less than 1\% (for the IV attack) to 37.8\% (for the BH attack). These findings imply that a much lower false alarm rate for benign traffic flows is guaranteed at the cost of a reduced detection rate for malicious flows as more nodes participate in the architecture, resulting in higher accuracy performance overall.

\begin{table}[!h]
\centering
\caption{Comparative TPR and FPR values obtained with two and nine participating nodes in the DCID architecture}
\begin{tabular}{llllll} \toprule
\multirow{2}{*}{\textbf{Attack}} & \multicolumn{2}{c}{\textbf{with two nodes}}                         & \textbf{} & \multicolumn{2}{c}{\textbf{with nine nodes}}                        \\ \cline{2-3} \cline{5-6}
                                 & \multicolumn{1}{c}{\textbf{TPR}} & \multicolumn{1}{c}{\textbf{FPR}} & \textbf{} & \multicolumn{1}{c}{\textbf{TPR}} & \multicolumn{1}{c}{\textbf{FPR}} \\ \midrule
\textbf{BH}                      & \cellcolor{gray25}0.908                            & 0.499                            &           & 0.892                            & \cellcolor{gray25}0.121                            \\
\textbf{SF}                      & \cellcolor{gray25}0.821                            & 0.633                            &           & 0.674                            & \cellcolor{gray25}0.281                            \\
\textbf{DI}                      & \cellcolor{gray25}0.864                            & 0.542                            &           & 0.774                            & \cellcolor{gray25}0.180                            \\
\textbf{DR}                      & \cellcolor{gray25}0.831                            & 0.602                            &           & 0.703                            & \cellcolor{gray25}0.237                            \\
\textbf{HF}                      & 0.932                            & 0.421                            &           & \cellcolor{gray25}0.940                            & \cellcolor{gray25}0.045                            \\
\textbf{IV}                      & \cellcolor{gray25}0.998                            & 0.004                            &           & 0.997                            & \cellcolor{gray25}0.000                            \\
\textbf{WP}                      & \cellcolor{gray25}0.876                            & 0.534                            &           & 0.820                            & \cellcolor{gray25}0.161   \\ \bottomrule                  
\end{tabular}
\label{tbl:tpr_fpr_dcid}
\end{table}

\section{The Proposed Method}
\label{section_federated-architecture}
Our analysis on architectures clearly demonstrates that the CIDwL exhibits the poorest detection performance, primarily due to its focus solely on local network traffic from the ID node. In contrast, CIDwG and DCID emerge as promising alternatives, as they capture a broader perspective by incorporating global network traffic through dedicated collaborative nodes. However, CIDwG and DCID compete with each other depending on the targeted attack (refer to Figure~\ref{fig:distributed_accr_number_ids}). Unlike CIDwL, both CIDwG and DCID introduce additional communication costs and energy consumption, highlighting a significant drawback, especially in LLNs. Our primary motivation in proposing FedID is to eliminate operation cost associated with the CIDwG and DCID architectures, while still maintaining detection performance at a level better than that of the CIDwL architecture. The proposed solution is discussed in detail in the following section.

\subsection{Federated Intrusion Detection (FedID)}
The proposed method, called federated intrusion detection (FedID), is based on federated learning~\cite{federated_learning} in which collaborators are present to develop and share their individual models to obtain a global aggregated model. Similar to CIDwG; FedID depends on multiple collaborator nodes (solely during the model construction phase), and they are responsible to build individual models based on their local network traffic as those in DCID. However, unlike CIDwG, collaborators in FedID are not required to transmit heavy traffic data to the root node. Instead, the weight parameters associated with their local models are aggregated to build a global model. In the proposed approach, the root node is responsible for developing a global detection model by aggregating incoming weight parameters that is then sent back to local trainers. This process allows collaborators to further improve their models for a better detection performance. The general steps of training FedID is given as follows, and this iterative process is repeated for a specified number of rounds, set to 100 in this study:
\begin{enumerate}[label=\roman*)]
    \item Initially, the root node constructs a global model randomly, which is then multicast to all collaborators.
    \item Collaborators, including the root node, download and update the incoming global model based on their respective local data. In this case, the eXtreme Gradient Boosting (XGBoost) algorithm~\cite{xgboost}, known for its compatibility in model aggregation in FL, is employed. For training, the same feature set~\cite{TL-GP} employed in the previous experiments is used.
    \item Collaborators then unicast their local model parameters to the root node. 
    \item Subsequently, the root node considers all local models for aggregation to train the global model. Here, the federated averaging (FedAvg)~\cite{federated_learning}, which averages the local gradient updates to form a global gradient update, is employed. Then, the global model becomes available for use by any node within the LLN. In this study, it is deployed at the root node. 
\end{enumerate}

It is crucial to note that the differences in local traffic patterns caused by the attacker's location lead to a non-IID data for FedID during model construction (i.e., training). To address this challenge, various approaches have been proposed in the literature, with global data sharing~\cite{data_sharing_fl} emerging as the most straightforward yet effective solution~\cite{fl_noniid_survey}. In this strategy, each collaborator only shares a small portion of local traffic samples with other collaborators to create a globally shared dataset. Hence, they train their local models using not just their local traffic data but also globally shared few traffic data provided by other collaborators. 

\begin{figure}[!h]
    \centering
    \includegraphics[width=\linewidth]{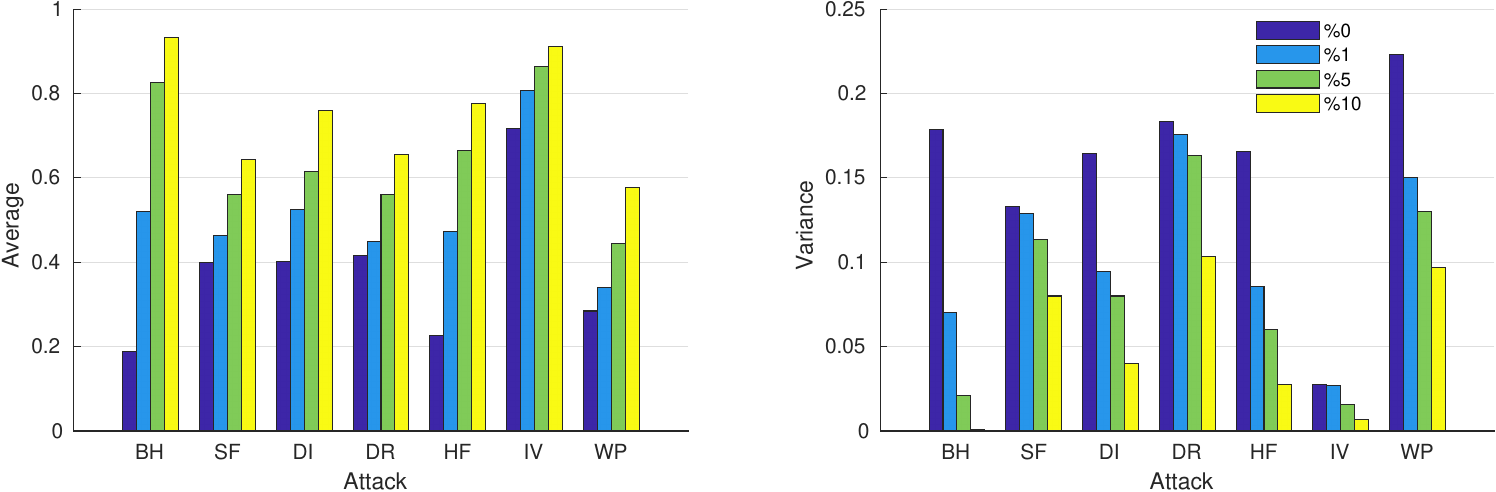}
    \caption{Divergence of local and global models as a function of data sharing ratio}
    \label{fig:cosine_sim}
\end{figure}

To demonstrate the presence of non-IID and its relation to data sharing, we conduct a preliminary experiment by calculating the divergence between local models and the global model, constructed through FedAvg, as a function of the data sharing ratio. Since models eventually converge to different parameters based on local data, weight divergence serves as a reliable indicator of non-IID across local datasets~\cite{fl_noniid_survey}. The average and variance of cosine similarity scores are shown comparatively for all attacks in Figure~\ref{fig:cosine_sim}.

Cosine similarity scores range from -1 to 1, with lower average scores indicating a more non-IID distribution and higher variance suggesting greater heterogeneity across local data. Therefore, with no data sharing (ratio = 0\%), we expect lower average scores and higher variance, confirming non-IID distribution among local network traffic. The figure shows that, except for Attack IV, local models display high diversity across all attack types at a 0\% share ratio, highlighting non-IID traits in local traffic. As the share ratio increases to 10\%, the divergence reduces considerably, and variance scores steadily decline with higher share ratios, particularly at 10\%, underscoring the role of data sharing in reducing local data heterogeneity.

Based on this observation, we design FedID such that $S$\% ($S \in \{0, 1, 5, 10\}$) of local traffic samples are shared among collaborators only during the model construction phase.

\subsection{Evaluation of FedID}
\subsubsection{Architectural comparison} In this section, we conduct a comparative evaluation of the architectures against seven targeted RPL attacks, assessing each architecture's detection performance as well as its communication and computation overhead, to demonstrate the applicability of FedID in LLNs.

\textbf{\textit{Effectiveness:}} FedID is thoroughly evaluated within the same network settings and attack types as used in the previous experiments. Then, it is compared with other architectures CIDwL, CIDwG, and DCID in order to highlight the primary contributions of FedID particularly in enhancing the detection performance and also reducing computational costs.

In testing of FedID, the global model aggregated at the root node is used. It is important to note that, unlike CIDwG and DCID, the model conducts intrusion detection autonomously without reliance on any collaborator or ID node during the testing step. In other words, the model analyzes and makes decision based on the local traffic of the root node.

\begin{figure}[!h]
    \centering
    \includegraphics[width=\linewidth]{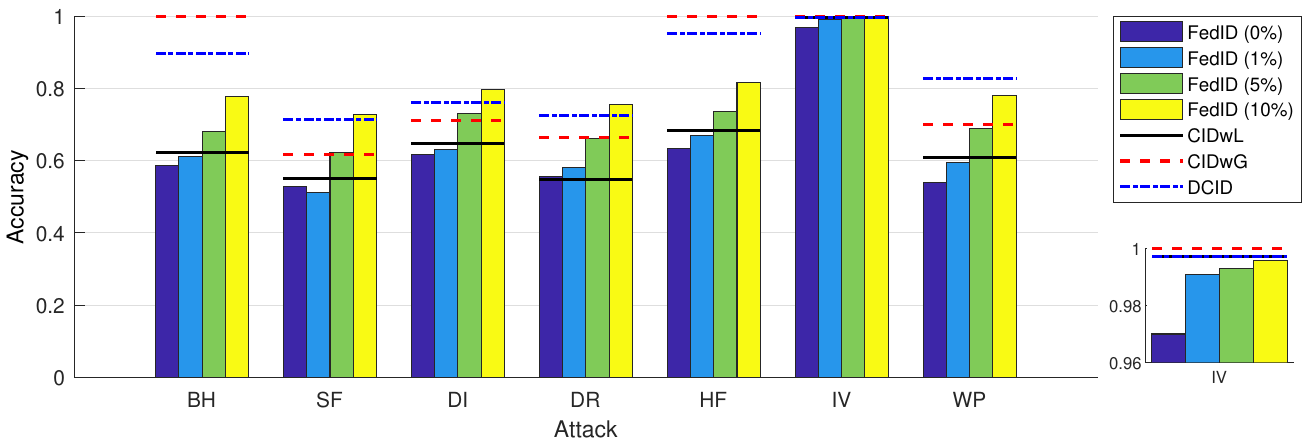}
    \caption{The comparison of the overall accuracy obtained by FedID and other architectures}
    \label{fig:FedID_CIDwG_single_attacker}
\end{figure}

Here, we consider the performance of CIDwG and DCID with nine collaborators (i.e., the nodes 0, 1, 10, 15, 8, 25, 14, 23, 24 in Figure~\ref{fig:topology}) that are the same as those in previous setting to ensure a fair comparison. In the experiments, we utilize Flower~\cite{flower}, a widely recognized FL framework, for implementing FL atop XGBoost. The default parameter settings for both FL and XGBoost, as described in~\cite{flower} (i.e., learning rate of 0.3 and maximum tree depth of 6), are employed in the experiments. Since the model operates independently at the root node in FedID in testing, we incorporate CIDwL with a configuration where the model is constructed and evaluated at the root node and include its detection performance for comparison here. The comparative accuracy results of the experiments are presented in Figure~\ref{fig:FedID_CIDwG_single_attacker}. It is important to emphasize again that these results represent the testing performance under conditions where the data and weight sharing mechanism takes place during training, but not in testing.

The experimental results show that FedID demonstrates increasing accuracy performance as the percentage of local traffic samples shared ($S$) increases during model construction for all attack types. Moreover, other than IV attack, FedID consistently outperforms the CIDwL architecture across all attack types when at least 5\% data is globally shared, underscoring the significance of incorporating local traffic data during model construction once again. Note that this observation is also supported by the performance gap between CIDwL and CIDwG; however, as compared to CIDwG, FedID is much more efficient due to significantly reduced traffic (collaborators share only 10\% of traffic samples during training at most, with no sharing during testing). This makes FedID more suitable for LLNs. An in-depth discussion on the efficiency is comparatively made in the following section.

In terms of accuracy comparison between CIDwG and FedID, even though collaborators no longer send local traffic data during the testing, FedID again demonstrates superior, or at least competitive, performance for SF, DI, DR, and WP attacks when it constructs global model with at least 5\% data sharing. CIDwG outperforms FedID only for BH and HF attacks, but this comes with the drawback of increased data size throughout the network during both training and testing. This can be attributed to the nature of these attacks, where local effects on the network are significant, and the continued transmission of traffic data by collaborators after model construction substantially aids the model in detecting these attacks in CIDwG.

DCID emerges as the most competitive architecture compared to CIDwL and CIDwG architectures. FedID outperforms DCID for SF, DI, and DR attacks with only 10\% data sharing during model construction. A better detection results could be obtained with higher data sharing ratio for BH, HF, and WP attacks. Like CIDwG, the decision of ID collaborators is periodically transmitted, contributing to increased overhead and reduced average network lifetime. This is why, particularly in testing, FedID appears more favorable than DCID in terms of efficiency. Additionally, the collaborators in DCID relies on local data to determine whether to raise an alarm, which is not always feasible for LLNs, particularly under high traffic loads. In contrast to DCID or CIDwG, the FedID model operates autonomously within a dedicated node in testing, eliminating the need for external knowledge to function.

The model learned by FedID is deployed at the root node, meaning only its local traffic contributes to the detection performance in testing. Since different regions of the network exhibit diverse traffic patterns within the attacker-involved network as shown earlier, transferring the final model to any other collaborator could prove beneficial and even essential to enhance its detection performance. To ascertain the full extent of FedID's capabilities, we upload the final model across all collaborators, restricting its operation solely to their respective local traffic data. As competitor architecture, we specifically selected CIDwL, as its architecture operates exclusively on individual nodes, focusing solely on their local traffic. It is important to note that to calculate the average of the best detection performances of CIDwL, we initially train separate models for each collaborator using their local traffic data, and subsequently test them on the same local traffic. This approach differs from FedID, where the model is solely tested on the collaborator nodes without training individual models for each. The average of best accuracy performances are comparatively shown for all attack types in Table~\ref{tbl:fedID_avg_best_accr}.
\begin{table}[!h]
\caption{The comparison of FedID across different data sharing ratios and CIDwL based on average best accuracy performance}
\centering
\begin{tabular}{cccccc}\toprule
\multirow{3}{*}{\textbf{Attack}} & \multicolumn{4}{c}{\textbf{FedID}}                 & \multirow{3}{*}{\textbf{CIDwL}} \\ \cline{2-5}
                                 & \multicolumn{4}{c}{\textbf{Data Sharing Ratio}}    &                                 \\  
                                 & \textbf{0\%} & \textbf{1\%} & \textbf{5\%} & \textbf{10\%} &                                 \\ \midrule
\textbf{BH}                      & 0.841      & 0.854      & 0.918      & 0.957       & \cellcolor{gray25}0.998                           \\
\textbf{SF}                      & 0.641      & 0.649      & \cellcolor{gray25}0.741      & \cellcolor{gray25}0.824       & 0.713                           \\
\textbf{DI}                      & 0.701      & 0.736      & 0.806      & \cellcolor{gray25}0.870       & 0.845                           \\
\textbf{DR}                      & 0.640      & 0.672      & \cellcolor{gray25}0.741      & \cellcolor{gray25}0.826       & 0.725                           \\
\textbf{HF}                      & 0.882      & 0.961      & 0.963      & 0.980       & \cellcolor{gray25}1.000                           \\
\textbf{IV}                      & \cellcolor{gray25}1.000      & \cellcolor{gray25}1.000      & \cellcolor{gray25}1.000      & \cellcolor{gray25}1.000       & \cellcolor{gray25}1.000                           \\
\textbf{WP}                      & 0.698      & 0.716      & 0.795      & \cellcolor{gray25}0.874       & 0.844 \\ \bottomrule
\end{tabular}
\label{tbl:fedID_avg_best_accr}
\end{table}

The findings in Table~\ref{tbl:fedID_avg_best_accr} indicate that FedID achieves higher detection accuracy as the data sharing ratio ($S$) increases, supporting the importance of data sharing in non-IID environments during model development. Compared to results obtained from the root node, there is a significant enhancement in detection performance when FedID's model operates on collaborator nodes, suggesting its effectiveness in distributed settings. This would also address the single point of failure in the network.

In comparison with the CIDwL architecture in Table~\ref{tbl:fedID_avg_best_accr}, FedID outperforms for SF, DI, DR, and WP attacks when at least 5\% of data is shared during model construction, while CIDwL shows slight superiority for BH and HF attacks. The promising detection performance of FedID for these attacks suggests potential further improvement with moderate increases in $S$. Moreover, in terms of operational cost, FedID proves more practical than CIDwL as it does not necessitate a separate model construction phase (i.e., training detection algorithm from scratch) before deployment on collaborator nodes. This aspect would facilitate the integration of new nodes into the network, especially considering the continuous inflow of new IoT devices with heterogeneous capabilities joining the system every day.

\textbf{\textit{Efficiency:}} In this section, we present both the communication and computational costs of each architecture and compare them.

The communication overhead (\( O_{\text{comm}} \)) of an architecture is calculated during both model training and testing using the formula below:
\begin{equation}
O_\text{comm} = N \times F \times S \times W   
\end{equation}
where \( N \), \( F \), \( S \), and \( W \) represent, respectively, the number of collaborators, the number of features, the size (in bytes, B) per feature, and the total number of windows representing the number of feature vectors shared between collaborators and the central node. In this study, we use nine collaborators (\( N = 9 \)) with each sending 35 features (\( F = 35 \)). For CIDwG and FedID, each feature takes 4 B (\( S = 4 \)), while for DCID, each alert takes 1 B (\( S = 1 \)). With 5-hour simulation at 60-second intervals, \( W \) is set to \( 300 \times \alpha \), where \( \alpha \) is the data-sharing ratio: 1.0 for CIDwG and DCID, and \( \alpha \in \{0.00, 0.01, 0.05, 0.10\} \) for FedID. The \( O_{\text{comm}} \) values of each architecture observed during training and testing are comparatively shown in Table~\ref{table:co}

\begin{table}[!h]

\scriptsize
\centering
\caption{Comparison of communication cost of architectures (in byte)}
\begin{tabular}{lccc}
\toprule
\textbf{Architecture}    & \textbf{Ratio ($\alpha$)} & \textbf{Training} & \textbf{Testing} \\ \toprule
\textbf{CIDwL}                  & N.A.      & 0 & 0                  \\ \midrule
\textbf{CIDwG}                  & 1.00      & 378$\times10^3$ & 378$\times10^3$                   \\ \midrule
\textbf{DCID}                   & 1.00      & 2700 & 2700                  \\ \midrule
\multirow{4}{*}{\textbf{FedID}} & 
                                 0.00      & 0 & \multirow{4}{*}{0} \\ 
                                & 0.01      & 3780&                    \\
                                & 0.05      & 189$\times10^2$ &                    \\
                                & 0.10      & 378$\times10^2$ & 
                              \\ \bottomrule 
\end{tabular}
\label{table:co}
\end{table}

As shown in the table, CIDwL is the most efficient (0 B) during training since it does not involve any data sharing with other nodes, followed by DCID (2.7 KB), FedID (up to 37.8 KB with 10\% sharing), and CIDwG (378 KB). In testing, however, both CIDwL and FedID remain lightweight at 0 B, as no data is shared, while DCID and CIDwG incur higher costs. In conclusion, when balancing communication cost and detection performance (refer to Figure~\ref{fig:FedID_CIDwG_single_attacker}), FedID (with 10\% sharing) emerges as the most promising option, despite the added communication cost during training. 

Unlike the other architectures, FedID requires parameter sharing exclusively during training. The communication cost from parameter sharing heavily depends on the depth of the XGBoost tree ($d$), with up to $2^d$ parameters exchanged for a limited number of training rounds. FedID shows a positive correlation between detection accuracy and $d$.

In addition to communication overhead, we also evaluate the architectures based on computational cost. Specifically, we consider the feature extraction cost ($\mathcal{O}(C_{\text{FE}})$) and the model cost ($\mathcal{O}(C_{\text{M}})$), which arise from the root node and the collaborator nodes. The model cost includes either the fitting cost ($\mathcal{O}(C_{\text{fit}})$) or the evaluation cost ($\mathcal{O}(C_{\text{eval}})$) during the training or testing phase. For simplicity, we denote the model cost as ($\mathcal{O}(C_{\text{M}})$) for both phases. All architectures have ($\mathcal{O}(C_{\text{FE}})$) and ($\mathcal{O}(C_{\text{M}})$).
Considering these, the total computational cost ($O_{\text{comp}}$) of the architecture is quantitatively expressed as:
\begin{equation}
    O_{\text{comp}} = N \times \mathcal{O}(C_{\text{FE}}) + M \times \mathcal{O}(C_{\text{M}}),
\end{equation}
where $N$ and $M$ denote the number of nodes responsible for feature extraction and model fitting/evaluation, respectively. Here, the collaborator nodes are set to the maximum ($=9$), with one collaborator node assigned at each level. The $O_{\text{comp}}$ performance of the architectures is comparatively presented in Table~\ref{table:comp}. 

\begin{table}[!h]
    \centering
    \caption{Comparison of computational cost of architectures}
    \tiny
    \begin{tabular}{lm{.5\linewidth}m{.2\linewidth}} \toprule
    \textbf{Architecture}    & \multicolumn{1}{c}{\textbf{Training}} & \multicolumn{1}{c}{\textbf{Testing}} \\ \toprule
     \textbf{CIDwL}    & \multicolumn{2}{c}{$\mathcal{O}(C_{\text{FE}}) + \mathcal{O}(C_{\text{M}})$}  \\ \midrule
     \textbf{CIDwG}    & \multicolumn{2}{c}{$9\times \mathcal{O}(C_{\text{FE}}) + \mathcal{O}(C_{\text{M}})$}  \\ \midrule
     \textbf{DCID}    & \multicolumn{2}{c}{$9\times \mathcal{O}(C_{\text{FE}}) + 9 \times \mathcal{O}(C_{\text{M}})$}  \\ \midrule
     \textbf{FedID}    & $8\times [\alpha^* \times \mathcal{O}(C_{\text{FE}})] + \mathcal{O}(C_{\text{FE}}) + \mathcal{O}(C_{\text{M}})$ & $\mathcal{O}(C_{\text{FE}}) + \mathcal{O}(C_{\text{M}})$\\ \bottomrule
     \multicolumn{3}{l}{\tiny $^*$data sharing ratio, $\alpha \in \{0.00, 0.01, 0.05, 0.10\}$}
    \end{tabular}
    
    \label{table:comp}
\end{table}

As shown in the table, CIDwL is the most computationally efficient architecture compared to the others, including FedID (except when $\alpha \neq 0$). For other sharing ratios, CIDwL is followed by FedID, which significantly lowers costs relative to CIDwG and DCID. In testing, however, both FedID and CIDwL exhibit minimal increases in computational cost, making them relatively efficient in comparison to the alternatives. While DCID show low communication cost, it is the most resource-intensive architecture in terms of computation cost. This is due to the additional costs incurred by the collaborator nodes, which are employed for both feature extraction and model fitting/evaluation. The main advantage of DCID lies in communication cost as expected. 

\subsubsection{Benchmark comparison on IoT dataset}
In addition to the comparative performance analysis of FedID on our custom RPL dataset, we have included the CICIoT2023 real-time benchmark dataset~\cite{ciciot2023}. The dataset features 33 IoT attacks classified into seven categories: DDoS, DoS, Recon, Web-based, Brute Force, Spoofing, and Mirai. We evaluated its detection capabilities by comparing it with recent ID solutions for IoT attacks. Among the competitors, multi-layer perceptron (MLP) is used in~\cite{abbas2023novel,NKORO2024101046}, and LSTM is used in~\cite{wang2023lightweight}. Note that the IDS proposed in~\cite{abbas2023novel} also based on FL.

The accuracy results are comparatively shown in Table~\ref{tbl:ciciot_comparison}. The table presents the detection performances of the algorithms based on binary and multi-class attack categories. \textit{Binary} in the table stands for binary classification task where the IDS discriminates between malicious attacks and benign traffic, while \textit{multi} stands for multi-class classification that refers to the identification of 34 categories (33 attack types and 1 category for benign traffic). Note that the output of the model is subject to logistic and softmax functions to handle binary- and multi-class classification tasks with FedID, respectively, and that the evaluations are made based on the accuracy metric. It is worth emphasizing that only `Recon' (ping sweep, OS scan, vulnerability scan, port scan, host discovery) and `Mirai' (GREIP flood, Greeth flood, UDPPlain) attack categories are targeted in~\cite{wang2023lightweight}. Therefore, its detection accuracy in the table is shown for these 8 attack types (marked with `$^*$').

\begin{table}[!h]
\centering
\caption{Comparison of accuracy performances on the CICIoT2023 dataset.}
\begin{tabular}{llcc}\toprule
\multirow{2}{*}{\textbf{Study (Year)}}       & \multirow{2}{*}{\textbf{Algorithm}} & \multicolumn{2}{c}{\textbf{Classification}} \\
&   & \textbf{Binary}   & \textbf{Multi}  \\ \midrule
Abbas et al.~\cite{abbas2023novel} (2023)                  &  FL with MLP                                & 99.00\%           & NA              \\
Nkoro et al.~\cite{NKORO2024101046} (2024)                  &      MLP                                      & 99.10\%           & 70.53\%         \\
Wang et al.~\cite{wang2023lightweight} (2023) & LSTM & NA                & 93.13\%$^*$         \\
\textbf{FedID}                               & FL with XGBoost                                                                                                         & \textbf{99.60\%}           & \textbf{98.80\%}        \\ \bottomrule
\multicolumn{4}{l}{NA: Not available.}\\
\end{tabular}
\label{tbl:ciciot_comparison}
\end{table}

As seen in the results, FedID demonstrates better performance than the other IDSs, excelling both in distinguishing malicious traffic from benign traffic (i.e., binary-class classification)  and in identifying specific attacks (i.e., multi-class classification). In comparison to its FL-based competitor~\cite{abbas2023novel}, FedID achieves slightly better detection accuracy due to the use of XGBoost. It is important to note that no data sharing occurs among nodes in this setup.

\section{Discussion}
\label{section_discussion}
In this section, the overall discussion of ID architectures is presented from different perspectives: effectiveness, cost and response time, threats to validity, privacy, and security.

\textit{\textbf{Effectiveness:}} The results clearly show that a central node is not enough to detect attacks at different locations. Hence, a distributed and cooperative architecture is more appropriate for RPL topologies. As the results suggest, at least one ID node could be placed at each level for effective detection. This is also important for some types of attacks, such as WP that have a higher impact locally. Although such architectures increase communication and computation cost, it is worth placing ID nodes at each level for reasonable accuracy and scalability. Moreover, a voting scheme could be preferred over sending features periodically for a lower communication cost, which also overcomes fragmentation issues in LLN. In the future, the features collected from the lower layer could be included for a better detection of other types of attack, namely BH and HF as in~\cite{canbalaban2020cross}.

The federated learning-based approach also shows promising results. By leveraging the decentralized nature of federated learning, data privacy is maintained while still allowing for the collaborative training of models across different nodes. Furthermore, partial data sharing among nodes could improve results even more, by providing a balance between data privacy and the richness of the shared information.

\textit{\textbf{Cost and response time:}} As pointed out earlier, although collaborative nodes enable detection systems to perform better than central-based architectures like CIDwL, they considerably increase network communication and computation costs. At this point, federated learning not only enhances the detection capabilities but also mitigates the need for extensive data transfer, further reducing communication overhead. Consequently, federated learning presents a viable solution for improving IDS performance in LLN environments without compromising scalability or accuracy. 

Distributed or cooperative architectures (CIDwG, DCID) not only introduce communication and computation costs but also delays in response time to attacks. Additionally, they may become impractical in environments with high packet loss, such as LLNs. Hence, a federated learning-based architecture like FedID could be a better solution for such networks, as it combines the detection capabilities of cooperative architectures with the prompt response time of centralized architectures. However, it is important to note that all these architectures, including CIDwL and FedID, require the IDS to wait for a certain period to gather sufficient traffic data before responding. This wait time is crucial for detection accuracy but can increase the network's response time to attacks. Therefore, it is essential to adjust the duration of this period to balance detection accuracy and prompt response based on the specific needs of the targeted IoT application. 

While FedID does not introduce any communication cost during testing, it requires both data and weight sharing during training. Data sharing is comparable to other models, as shown in Table~\ref{table:co}, but FedID additionally involves weight sharing. In this study, we employ the XGBoost algorithm, whose accuracy is positively correlated with tree depth. However, this improvement comes at the cost of increased weight sharing, highlighting a trade-off between detection performance and communication overhead during training. Similarly, while it incurs higher computational costs than CIDwL during training, FedID proves to be one of the more efficient architectures in terms of computation cost during testing.  

\textit{\textbf{Threats to validity:}} As shown in the results, attackers' locations have a clear impact on their detectability. Hence, if attackers know the placement of IDs, they could evade detection by performing their attacks far from ID nodes. Moreover, mobile attackers could exploit this knowledge by adjusting their position dynamically, further reducing their chances of being detected by avoiding areas with strong intrusion detection coverage. Therefore, placing an ID node at each level becomes more important, as suggested in our findings. A federated learning-based IDS, where each node participates with its local updates, offers a promising alternative. This approach allows each node to contribute to a global model, enhancing privacy and security. By aggregating local updates, the system can achieve high detection accuracy across various network segments, making it more difficult for attackers to evade detection regardless of their location.

A central ID node is one of the main drawbacks of all architectures. While FL mitigates some risks associated with a single point of failure by decentralizing the training process, there is still a potential single point of failure when aggregating the global model updates and running the global model after training. For such single point of failure points, there should always be another backup node that could work in case of a failure in the main node. Another solution could be to place the central ID node in the cloud, outside the network. Another option is to collect alarms at each ID node in DCID or to generate global models at more than one node in FedID. Even though this increases communication cost, it is fairly acceptable for short messages.

\textit{\textbf{Privacy:}} While CIDwG produces higher detection accuracy, it requires data aggregation at a singular point. This aggregation exposes devices, undermining the privacy they aim to protect. Therefore, DCID, which shares local alarms, or FedID could be better alternatives. FL decentralizes the learning process, distributing it across multiple nodes within the network. This also ensures data privacy, as information no longer needs to traverse the entire network to a central repository for processing after training. Nevertheless, FL encounters its own set of challenges. One significant challenge is the necessity for substantial data volumes at each node to construct a high-performance model \cite{9060868, 9464278}. In the context of IoT, where devices frequently operate under severe resource constraints, this requirement poses a significant risk. A potential mitigation strategy, as proposed by \cite{8770530}, entails leveraging trusted neighboring nodes for data sharing or model training, thereby augmenting the overall effectiveness of the FL approach without unduly burdening individual devices. This strategy, applied in this study, underscores the delicate balance between effectiveness and privacy in deploying FL for IoT intrusion detection.

\textit{\textbf{Security:}} 
In all architectures, intrusion detection models are automatically generated using machine learning techniques. However, this makes them vulnerable to adversarial attacks, including federated learning due to its decentralized framework \cite{10274102, NABAVIRAZAVI202428}. Additionally, as discussed earlier, a central aggregation point exposes such single points to denial-of-service (DoS) attacks.     

\section{Conclusion}
\label{section_conclusion}
While many advancements in RPL security focus on analyzing attacks and developing detection methods, the placement of intrusion detection nodes is crucial as attacks can originate from various locations. In this study, we addressed this research problem through extensive simulations, exploring three intrusion detection architectures with considerations for performance, communication cost, and security.

The experimental results demonstrate the effectiveness of distributed and collaborative architectures in detecting attacks from various locations. However, these architectures significantly increase communication and computation costs, resulting in detection delays and privacy violations. Moreover, the communication among ID nodes is prone to drops in lossy networks. Therefore, we propose a new approach based on FL, which is shown to be effective in terms of effectiveness, communication and computation cost (particularly in testing), response time, and privacy. Consequently, our proposed approach, FedID, emerges as a suitable solution for LLNs.

We believe that our study provides valuable guidelines for researchers developing suitable intrusion detection techniques for RPL, emphasizing the importance of IDS locations with low communication costs.

\section*{Acknowledgment}
This study has been supported by the ADA project (Project ID: C2023/1-4) under the EUREKA Cluster CELTIC-NEXT program and TÜBİTAK-TEYDEB Grant Program (Project No: 9230039).

\bibliographystyle{IEEEtran}
\bibliography{manuscript}

\end{document}